# Row-Sparse Discriminative Deep Dictionary Learning for Hyperspectral Image Classification

Vanika Singhal, and Angshul Majumdar, *Senior Member, IEEE*

*Abstract*—In recent studies in hyperspectral imaging, biometrics and energy analytics, the framework of deep dictionary learning has shown promise. Deep dictionary learning outperforms other traditional deep learning tools when training data is limited; therefore hyperspectral imaging is one such example that benefits from this framework. Most of the prior studies were based on the unsupervised formulation; and in all cases, the training algorithm was greedy and hence sub-optimal. This is the first work that shows how to learn the deep dictionary learning problem in a joint fashion. Moreover, we propose a new discriminative penalty to the said framework. The third contribution of this work is showing how to incorporate stochastic regularization techniques into the deep dictionary learning framework. Experimental results on hyperspectral image classification shows that the proposed technique excels over all state-of-the-art deep and shallow (traditional) learning based methods published in recent times.

## I. INTRODUCTION

IN the recent past, deep learning has been gaining popularity in hyperspectral image classification [1-14]. All the standard deep learning models – Deep Belief Network (DBN) [1-3], Stacked Autoencoder (SAE) [4, 5], Deep Recurrent Neural Network (DRNN) [6], variants of PCA-Net [7, 8] and Convolutional Neural Network [9-14] have been employed in this context.

For successful performance, deep learning requires a large volume of training data. Unfortunately, this is impractical for hyperspectral image classification. Therefore, several modifications need to be made to the classical deep learning architectures to fit the said problem. In a recent work [15] a novel active learning methodology has been proposed to address the issue of limited training data.

In a recent study [16] a new deep learning tool called deep dictionary learning (DDL) has been proposed. The basic idea there in, is to use dictionary learning as basic building blocks for a deeper architecture. Intuitively speaking, the coefficients / feature from one layer of dictionary learning acts as an input to the subsequent layer leading to a deep network. It was shown in [16] that DDL can operate within the limited training data regime and yet yield a performance superior to standard deep learning tools like DBN, SAE and CNN.

Given the success of DDL in hyperspectral classification [16] amongst other areas like biometrics [17], energy analytics [18] and benchmark deep learning problems [19], we propose to build a new classifier based on the DDL framework. All prior studies on deep dictionary learning follow the greedy training paradigm; and hence are sub-optimal. Since each of the layers are learnt separately, the shallower layers influence the deeper ones, but not vice versa. For optimal training, all the layers should be learnt jointly. Besides, [17-19] are all unsupervised learning approaches – they can only extract the features but require an off-the-shelf classifier for classification. Only [16] is a supervised classification technique; albeit a suboptimal (greedy) one.

There are a few major contributions of this work. For the first time we show how the deep dictionary learning framework can be solved in a joint fashion. All the layers are updated jointly along with the final level of coefficients. The second contribution of this work is in proposing a new discriminative penalty. The third contribution of this work is to show how stochastic regularization techniques can be incorporated into the deep dictionary learning framework. This has not been attempted before.

The rest of the work will be organized into several sections. Relevant background will be discussed in section II. The proposed technique will be described in section III. The experimental results are detailed in section IV. Finally, the conclusions of this work and future direction of research will be discussed in section V.

## II. LITERATURE REVIEW

### A. Deep Learning in Hyperspectral Classification

In image analysis and computer vision problems CNN is the most popular choice. However, CNNs are data hungry and fail to perform in limited training data regime. This is the reason why other deep learning techniques like DBN [1-3] continue to be popular. In [1], a basic DBN is used with logistic regression based classification. The formulation is made supervised in [2] by using group-sparsity – a technique proposed in [20]. In [3], a diversifying pre-training technique is used for improving the performance of DBN based deep neural network.

The problem of DBN is that, it can only handle limited variability in the inputs; they need to be between 0 and 1. Some kind of batch normalization is required to bring all inputs within this range. However, given that the inputs (for

V. Singhal and A. Majumdar, are with Indraprastha Institute of Information Technology, New Delhi, 110020, India (email: vanikas@iiitd.ac.in and angshul@iiitd.ac.in) .

imaging problems) are spatially correlated, such arbitrary change in input values may decimate the local correlation information and reduce overall performance.

Much like DBN, SAE's have been used for hyperspectral image classification. The work [4] uses a standard stacked autoencoder for classification using logistic regression. In [5], some extra penalties are added to the basic stacked autoencoder formulation in order to incorporate contextual information.

Unlike DBNs stacked autoencoders can handle arbitrary input values. But the problem with SAE's is that given the same volume of data, the number of parameters it needs to learn is twice that of DBN (equal number of encoders and decoders). Owing to this, SAEs are likely to suffer from over-fitting.

Recurrent Neural Network (RNN) and its variants (echo state network, long short-term memory network) are applicable to ordinal data, for example time series data. In recent times, high impact conferences like CVPR and ICCV are publishing papers on image analysis and computer vision using RNNs (and its variants). In the same spirit, [6] used RNN and [7] used LSTM for analyzing hyperspectral images.

It must be noted that there is no theoretical justification to apply RNNs for image analysis. Such models are only applicable to ordered data (such as time series). Pixels are not ordered. RNNs require arbitrary ordering in order to justify its usage. This arbitrary ordering may affect the overall performance.

The concept of PCA-Net [21] has been proposed recently. The framework is similar to CNN. Each layer constitutes of a filtering and pooling operation. But instead of convolving, the image is filtered using principal component analysis. The study [8] proposes a non-linear variant of PCA-Net. In [9] instead of using PCA, they use vertex component analysis into the PCA-Net framework for hyperspectral image analysis.

All the studies on CNN based analysis [10-14] follow the same basic framework. They only differ from each other, in the choice of the number of layers, size and number of convolutional kernels and type of pooling used. It must be remembered that CNNs are data hungry, so studies in hyperspectral image classification propose slight modifications to suit the goal.

Owing the problem of data scarcity in deep learning, an interesting proposal has been laid in [15]. They propose an active learning methodology to choose new training samples in a recursive fashion. However, active learning requires constant feedback from the user; this does not seem to be very practical given the scenario. It is unlikely that such frameworks that require constants feedbacks will be pragmatic and popular.

It has been seen in [16] that the newly proposed framework of deep dictionary learning outperforms many traditional deep learning tools in hyperspectral image analysis. In other areas [17-19] a similar trend has been envisaged. The DDL framework excels when there is limited training data is available.

### B. Deep Dictionary Learning

Since the concept of deep dictionary learning is relatively new, we believe that a brief introduction on this topic will help readers and make the paper self-contained. Let us reiterate the basic idea of deep dictionary learning. In (shallow) dictionary learning, one level of dictionary is used to represent the signal. In DDL, multiple layers of dictionaries are used to represent the signal.

Formally, in dictionary learning, the training set ($X$) is decomposed into a dictionary ($D$) and a matrix of coefficients ($Z$). This is represented as,

$$X = DZ \qquad (1)$$

Here the training signals are stacked as columns of $X$; the corresponding features are in $Z$. Usually the learning is expressed as,

$$\min_{D,Z} \|X - DZ\|_F^2 + \lambda \|Z\|_0 \qquad (2)$$

The first term is for data fidelity; the $l_0$-norm imposes sparsity on the coefficients. There are well known algorithms to solve the dictionary learning problem (2).

Dictionary learning basically factors the matrix into the dictionary matrix and a coefficient matrix. In recent years, following the success of deep learning, several studies have proposed deep matrix factorization [22, 23]; here the data is decomposed into multiple layers of dictionaries ($D_1$, $D_2$, $D_3$, …) and one final layer of coefficients ($Z$). Mathematically this is represented as follows (shown for three layers),

$$X = D_1 D_2 D_3 Z \qquad (3)$$

The problem with this formulation is that, there is no activation function between the layers; there is a possibility that the multiple layers of dictionaries collapse into a single layer.

This issue is rectified in deep dictionary learning [19]; in there an activation function is incorporated between each layer. For a three layer DDL, the formulation is given by,

$$X = D_1 \varphi(D_2 \varphi(D_3 Z)) \qquad (4)$$

Usually a non-linear activation function like sigmoid or tanh is used.

The exact solution for DDL should have been,

$$\min_{D_1, D_2, D_3, Z} \|X - D_1 \varphi(D_2 \varphi(D_3 Z))\|_F^2 + \lambda \|Z\|_0 \qquad (5)$$

However, it must be noted that none of the prior studies [16-19] solves (5) directly. All of them solve it greedily – one layer at a time.

In greedy learning, for the first layer, one substitutes $Z_1 = \varphi(D_2 \varphi(D_3 Z))$. Therefore, (5) boils to the following problem

$$\min_{D_1, Z_1} \|X - D_1 Z_1\|_F^2 \qquad (6)$$

This is a standard matrix factorization problem.

Once the coefficients from the first layer is learnt, it is input to the second layer. Here one substitutes: $Z_2 = \varphi(D_3 Z)$. This leads to –

$$Z_1 = \varphi(D_2 Z_2) \qquad (7)$$

Equivalently this can be expressed as,

$$\varphi^{-1}(Z_1) = D_2 Z_2$$

Since the activation functions are unitary, inverting them is trivial. It is easy to solve (7) via matrix factorization.

$$\min_{D_2, Z_2} \left\| \varphi^{-1}(Z_1) - D_2 Z_2 \right\|_F^2 \qquad (8)$$

For the last layer, we have,

$$Z_2 = \varphi(D_3 Z) \\ \equiv \varphi^{-1}(Z_2) = D_3 Z \qquad (9)$$

Here the same principle as before has been used to invert the activation function and represent the last layer in the equivalent form.

Since the coefficients in the last layer are supposed to be sparse, one solves –

$$\min_{D_3, Z} \left\| \varphi^{-1}(Z_2) - D_3 Z \right\|_F^2 + \lambda \|Z\|_0 \qquad (10)$$

This is the formulation used in [17-19].

This concludes the training phase. In testing, the learnt dictionaries are used to generate the test feature. This is expressed as,

$$\min_{z_{test}} \left\| x_{test} - D_1 \varphi(D_2 \varphi(D_3 z_{test})) \right\|_F^2 + \lambda \|z_{test}\|_0 \qquad (11)$$

Using substitutions like before, this (11) too is solved greedily.

Building on this greedy framework, a supervised solution was proposed for hyper-spectral image classification [16]. After the final layer of greedy deep dictionary learning, a label consistency layer [34] was added, leading to the following formulation –

$$\min_{D_1, D_2, D_3, Z} \left\| X - D_1 \varphi(D_2 \varphi(D_3 Z)) \right\|_F^2 + \lambda \|Z\|_0 + \|T - MZ\|_F^2 \qquad (12)$$

Here $T$ are the target labels and $M$ the linear classifier map from the features (from the deepest layer) to the targets. This is a supervision term borrowed from [35].

However, note that [16] does not solve (12) jointly. Each layer of deep dictionary learning (till the penultimate layer) is solved piecemeal via a standard algorithm like KSVD; the final layer uses the implementation from [35].

## III. PROPOSED APPROACH

The problem with the greedy training paradigm in deep learning is that it is sub-optimal. The shallower layers influence the deeper layer but not vice versa. This is the first work that proposes to learn all the layers of deep dictionary learning in joint fashion.

Most prior studies on DDL were unsupervised. It is well known in machine learning literature, and specifically in deep learning that discriminative learning schemes indeed improve classification performance. This has been seen for both autoencoders [24, 25] and deep belief networks [26, 27]. This motivates our work to incorporate such penalties into the deep learning framework.

Our proposed work is different to [16] in two major ways. First, our work optimally solves all the variables in a joint fashion. Second, it uses a more sophisticated supervision penalty than label consistency.

### A. Formulation

Our formulation stems from the basic definition of 'discrimination' – samples within the same class will be close to each other and samples between classes will be far from each other.

The basic structure of deep dictionary learning remains the same as before; we would like to solve (5). However, since we are incorporating supervision / discrimination into the features we propose modifications on top of the basic formulation. The first modification we propose is to impose class-sparsity, i.e. instead of allowing features from the same class to be arbitrarily sparse we impose structure. We postulate that the features from the same class will have the same support, i.e. will have the same sparsity pattern. In other words, the positions of the non-zero coefficients for features of the same class will be the same for all samples. This has been introduced into the DBN framework in [20] and in the autoencoder framework in [28]. This modification is represented as follows –

$$\min_{D_1, D_2, D_3, Z} \left\| X - D_1 \varphi(D_2 \varphi(D_3 Z)) \right\|_F^2 + \lambda \sum_c \|Z_c\|_{2,0} \qquad (13)$$

Here the $l_{2,0}$-norm is defined as the number of non-zero rows of the matrix. This imposes row-sparsity. The sub-script 'c' denotes the class and $Z=[Z_1/…/Z_c/…]$. Therefore the cost function imposes supervised class sparsity on each class 'c'.

This cost is supervised but does not make the features discriminative. Since we are not imposing distinction between the classes; there is no way we can guarantee that the features from different classes will not have the same support. To impose discrimination, we need to ensure that the support between features of two classes are different. This is ensured by the second penalty.

$$\min_{D_1, D_2, D_3, Z} \left\| X - D_1 \varphi(D_2 \varphi(D_3 Z)) \right\|_F^2 + \lambda \sum_c \|Z_c\|_{2,0} \\ - \mu \sum_{k \neq c} \left\| \overline{Z}_k - Z_c \right\|_0 \qquad (14)$$

Here $\overline{Z}_k$ is the mean of the $k^{th}$ class repeated the same number of times as the number of samples in class c. This additional term ensures that the support between two different classes (k and c) are as distinct as possible; since we are maximizing the $l_0$-distance between the classes we need the negative sign.

Note that this is a completely different formulation from the label consistency penalty used in [16]; this new discrimination term is a major contribution of this paper.

This concludes the formulation for training. For testing, the goal is to assign a class label for the test sample $x_{test}$. For that, we need to generate the corresponding feature. This is achieved by solving the same problem as in the greedy case (11). This is repeated here for the sake of convenience.

$$\min_{z_{test}} \left\| x_{test} - D_1 \varphi(D_2 \varphi(D_3 z_{test})) \right\|_F^2 + \lambda \|z_{test}\|_0$$

Once the sparse feature is obtained, it is compared with the training samples for class assignment. In this work, the assignment is done in two ways.

In the first approach, once the training is complete, the exact training features bears no importance, only the sparse binary support of each class is required. An example may clarify the concept better.

| samples #1 | | | Binary Rep |
|---|---|---|---|
| 0 | 0 | 0 | 0 |
| 0 | 0 | 0 | 0 |
| 0.5 | 0.7 | 0.4 | 1 |
| 0 | 0 | 0 | 0 |
| 0.3 | 0.2 | 0.2 | 1 |
| 0 | 0 | 0 | 0 |

| samples #2 | | | | Binary Rep |
|---|---|---|---|---|
| 1.1 | 0.5 | 0.9 | 1.2 | 1 |
| 0 | 0 | 0 | 0 | 0 |
| 0 | 0 | 0 | 0 | 0 |
| 0.1 | 0.6 | 0.4 | 0.5 | 1 |
| 0 | 0 | 0 | 0 | 0 |
| 0.2 | 0.4 | 0.4 | 0.2 | 1 |

**Fig. 1.** Left – Sparse Features from Class 1 and its binary representation; Right – Features from Class 1 and its binary representation

Consider a toy problem where the sparse features for classes 1 and 2 are shown in Fig. 1. Class 1 has 3 samples and class 2 has 4. Virtually, instead of storing the features after the training phase we can just store the binary representation vector for each class. Once the test feature is generated, we generate its corresponding binary representation. The binary representation of the test data is compared with the binary representation of each sample via a binary dot product / Hamming distance. In practice, it boils down to the computation of the $l_0$-norm between the features of the training samples and the test sample; note that the $l_0$-norm is computed directly on the features and not on the binary representation. This is very efficient with linear complexity. The test sample is assigned to the class having the maximum similarity / minimum distance.

There may be an issue with the aforesaid approach. Assume that (for a problem different from the aforesaid toy problem), the binary representation of class 1 is [0,0,1,1,0,0], and the binary representation of class 2 is [0,0,1,0,1,0]. The binary representation of the test sample is [0,0,1,1,1,0]. For such a sample, the Hamming distance from both the classes 1 and 2 will be the same, and we cannot assign a unique class. In order to resolve this issue, in the second approach, instead of computing the $l_0$-norm we compute the $l_1$-norm between the features of the training and test samples; since they account for the magnitude and not only the position, such ties can be avoided. We assign the test sample to the class having the minimum absolute distance.

*B. Derivation*

For training we need to solve (14). We following the Split Bregman technique [29, 30]. We introduce a proxy $Z^1 = \varphi(D_2\varphi(D_3 Z))$. This leads to the following expression,

$$\min_{D_1,D_2,D_3,Z,Z^1} \|X - D_1 Z^1\|_F^2 + \lambda \sum_c \|Z_c\|_{2,0} - \mu \sum_{k \neq c} \|\bar{Z}_k - Z_c\|_0 \quad (15)$$
$$\text{s.t. } Z^1 = \varphi(D_2\varphi(D_3 Z))$$

Note that $Z^1$ basically corresponds to the features from the first level of dictionary learning.

In the second step, we introduce another proxy variable $Z^2 = \varphi(D_3 Z)$. This leads to,

$$\min_{D_1,D_2,D_3,Z,Z^1,Z^2} \|X - D_1 Z^1\|_F^2 + \lambda \sum_c \|Z_c\|_{2,0} - \mu \sum_{k \neq c} \|\bar{Z}_k - Z_c\|_0 \quad (16)$$
$$\text{s.t. } Z^1 = \varphi(D_2 Z_2), Z^2 = \varphi(D_3 Z)$$

The augmented Lagrangian is formulated from (16) by incorporating the Bregman relaxation variable ($B_1$ and $B_2$) for each proxy.

We form the augmented Lagrangian –

$$\min_{D_1,D_2,D_3,Z,Z^1,Z^2} \|X - D_1 Z^1\|_F^2 + \lambda \sum_c \|Z_c\|_{2,0} - \mu \sum_{k \neq c} \|\bar{Z}_k - Z_c\|_0$$
$$+\eta_1 \|Z^1 - \varphi(D_2 Z_2) - B_1\|_F^2 + \eta_2 \|Z^2 - \varphi(D_3 Z) - B_2\|_F^2 \quad (17)$$

Here $B_1$ and $B_2$ are the relaxation variables.

The final problem (17) is solved by the alternating direction method of multipliers (ADMM) [29]; the formulation (17) is segregated into the following sub-problems.

P1: $\min_{D_1} \|X - D_1 Z^1\|_F^2$

P2: $\min_{D_2} \|Z^1 - \varphi(D_2 Z^2) - B_1\|_F^2 \equiv \min_{D_2} \|\varphi^{-1}(Z^1 - B_1) - D_2 Z^2\|_F^2$

P3: $\min_{D_3} \|Z^2 - \varphi(D_3 Z) - B_2\|_F^2 \equiv \min_{D_3} \|\varphi^{-1}(Z^2 - B_2) - D_3 Z\|_F^2$

P4: $\min_{Z^1} \|X - D_1 Z^1\|_F^2 + \eta_1 \|Z^1 - \varphi(D_2 Z^1) - B_1\|_F^2$

P5: $\min_{Z^2} \eta_1 \|Z^1 - \varphi(D_2 Z^2) - B_1\|_F^2 + \eta_2 \|Z^2 - \varphi(D_3 Z) - B_2\|_F^2$
$\equiv \min_{Z^2} \eta_1 \|\varphi^{-1}(Z^1 - B_1) - D_2 Z_2\|_F^2 + \eta_2 \|Z^2 - \varphi(D_3 Z) - B_2\|_F^2$

P6: $\min_Z \eta_2 \|Z^2 - \varphi(D_3 Z) - B_2\|_F^2 + \lambda \sum_c \|Z_c\|_{2,0} - \mu \sum_{k \neq c} \|\bar{Z}_k - Z_c\|_0$
$\equiv \min_Z \eta_2 \|\varphi(Z^2 - B_2) - D_3 Z\|_F^2 + \lambda \sum_c \|Z_c\|_{2,0} - \mu \sum_{k \neq c} \|\bar{Z}_k - Z_c\|_0$

For sub-problems P2, P3, P4 and P5 the equivalent form is trivial to obtain since the non-linear activation functions are unitary and invertible – a property used by all prior works on DDL. All the sub-problems except P6 are easy to solve, since they are simple least squares problems with analytic solutions in the form of Moore Penrose Pseudoinverse.

Sub-problem P6 is slightly more involved since it does not have a direct solution. Although not exactly separable, we can decouple the problem and solve for each $Z_c$. This leads to,

$$\min_{Z_c} \eta_2 \|\varphi(Z_c^2 - B_2) - D_3 Z_c\|_F^2 + \lambda \|Z_c\|_{2,0} - \mu \|\bar{Z}_k - Z_c\|_0 \quad (18)$$

To solve (18) we need to invoke the Split Bregman technique once more. We introduce a proxy variable $P = Z_k - Z_c$. The corresponding augmented Lagrangian after Bregman relaxation is expressed as,

$$\min_{Z_c,P} \eta_2 \|\varphi(Z_c^2 - B_2) - D_3 Z_c\|_F^2 + \lambda \|Z_c\|_{2,0} - \mu \|P\|_0$$
$$+\gamma \|P - (\bar{Z}_k - Z_c) - C\|_F^2 \quad (19)$$

Here $C$ is the relaxation variable. Using ADMM once again, (19) can be split into the following two sub-problems.

S1: $\min_{Z_c} \eta_2 \|\varphi(Z_c^2 - B_2) - D_3 Z_c\|_F^2 + \lambda \|Z_c\|_{2,0}$
$+\gamma \|P - (Z_k - Z_c) - C\|_F^2$

S2: $\min_P \gamma \|P - (Z_k - Z_c) - C\|_F^2 - \mu \|P\|_0$

Sub-problem S1 is a solved problem. It can be solved by Simultaneous Orthogonal Matching Pursuit. Sub-problem S2

is not so trivial; instead of minimizing the $l_0$-norm (as is done in all standard problems in optimization, signal processing and machine learning) here we have to maximize it. However, the closed form solution readily presents itself from the proximal operator for S2. It is,

$$P = \begin{cases} Z_k - Z_c + C & \text{if } \frac{\mu}{2\gamma} < |\bar{Z}_k - Z_c + C| \\ \mu/2\gamma & \text{otherwise} \end{cases} \quad (20)$$

This concludes the derivation of sub-problem P6. The final step is to update the Bregman relaxation variables. This is achieved by simple gradient decent.

$$B_1 \leftarrow Z^1 - \varphi(D_2 Z_2) - B_1 \quad (21a)$$
$$B_2 \leftarrow Z^2 - \varphi(D_3 Z) - B_2 \quad (21b)$$
$$C \leftarrow P - (Z_k - Z_c) - C \quad (21c)$$

This concludes the training algorithm.

For testing one needs to solve (11); repeated here for the sake of convenience.

$$\min_{z_{test}} \left\| x_{test} - D_1 \varphi(D_2 \varphi(D_3 z_{test})) \right\|_F^2 + \lambda \left\| z_{test} \right\|_0$$

We use similar kind of proxies as before: $z_{test}^1 = \varphi(D_2 \varphi(D_3 z_{test}))$ for the first level and $z_{test}^2 = \varphi(D_3 z_{test})$ for the second. This leads to the following augmented Lagrangian formulation after introducing the Bregman proxy variables $b_1$ and $b_2$.

$$\min_{z_{test}, z_{test}^1, z_{test}^2} \left\| x_{test} - D_1 z_{test}^1 \right\|_2^2 + \lambda \left\| z_{test} \right\|_0$$
$$+ \eta_1 \left\| z_{test}^1 - \varphi(D_2 z_{test}^2) - b_1 \right\|_2^2 + \eta_2 \left\| z_{test}^2 - \varphi(D_3 z_{test}) - b_2 \right\|_2^2 \quad (22)$$

Following ADMM, (22) can be split into the following sub-problems.

$$\text{T1:} \min_{z_{test}} \eta_2 \left\| z_{test}^2 - \varphi(D_3 z_{test}) - b_2 \right\|_2^2 + \lambda \left\| z_{test} \right\|_0$$
$$\equiv \eta_2 \left\| \varphi^{-1}(z_{test}^2 - b_2) - D_3 z_{test} \right\|_2^2 + \lambda \left\| z_{test} \right\|_0$$
$$\text{T2:} \min_{z_{test}^1} \left\| x_{test} - D_1 z_{test}^1 \right\|_2^2 + \eta_1 \left\| z_{test}^1 - \varphi(D_2 z_{test}^2) - b_1 \right\|_2^2$$
$$\text{T3:} \min_{z_{test}^2} \eta_1 \left\| z_{test}^1 - \varphi(D_2 z_{test}^2) - b_1 \right\|_2^2 + \eta_2 \left\| z_{test}^2 - \varphi(D_3 z_{test}) - b_2 \right\|_2^2$$
$$\equiv \min_{z_{test}^2} \eta_1 \left\| \varphi^{-1}(z_{test}^1 - b_1) - D_2 z_{test}^2 \right\|_2^2 + \eta_2 \left\| z_{test}^2 - \varphi(D_3 z_{test}) - b_2 \right\|_2^2$$

Sub-problem T1 (in the equivalent form) is a simple $l_0$-minimization problem that can be either solved via orthogonal matching pursuit or iterative hard thresholding. Sub-problem T2 and T3 (in equivalent form) are simple least squares minimization problems which can be solved via pseduoinverses. As in the training phase, we update the Bregman relaxation variables by simple gradient descent.

$$b_1 \leftarrow z_{test}^1 - \varphi(D_2 z_{test}^2) - b_1 \quad (23a)$$
$$b_2 \leftarrow z_{test}^2 - \varphi(D_3 z_{test}) - b_2 \quad (23b)$$

This concludes the derivation of the testing algorithm.

Both the training and the testing algorithms require solving pseudoinverses and iterative thresholding. The complexity of pseudoinverse is $O(N^3)$; $N$ is the number of elements in the matrix. The complexity (per iteration) of thresholding is $O(N^2)$ (for matrix vector products) + $O(N)$ for thresholding. But about $O(N)$ steps are needed for convergence. Hence the total complexity of the thresholding sub-problems are approximately $O(N^3)$. Therefore, the overall complexity of our training and testing algorithms are $O(N^3)$.

### C. Stochastic Regularization

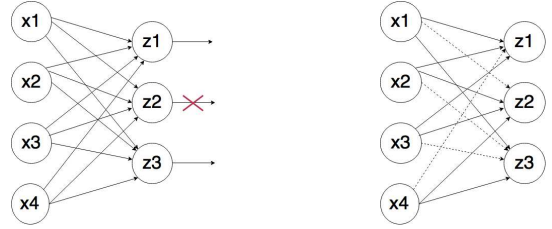

**Fig. 2.** (a) Left – DropOut Regularization. (b) Right – DropConnect[1]

There are two stochastic regularization techniques that are popular in deep learning. They are DropOut [31] and DropConnect [32]. These regularization techniques have been used for traditional neural networks, but never for dictionary learning. This is the first work that shows how such stochastic regularization techniques can be incorporated in deep dictionary learning.

The main idea in DropOut is to randomly drop units (along with their connections) from the network during training. This prevents units from co-adapting too much. Suppose we have training data $X$; in every iteration of DropOut some randomly chosen output units along with their connection weights are set to zero, as shown in Fig. 2. Here, out of three output neutrons, z2 (selected randomly) is dropped.

DropConnect is a generalization of DropOut. Here a set of randomly selected connections of network are set to zero. It works similar to DropOut regularization technique, except that, instead of dropping the whole unit, some connections are dropped. This makes output units partially active, as shown in Fig. 2. The dotted lines here show the dropped connections.

Both of them are regularization techniques. Given that we are interested in addressing the hyperspectral image classification problem, where training data is always parsimonious incorporating these into the deep dictionary learning framework to prevent over-fitting may prove beneficial.

We can incorporate DropOut type regularization, by randomly imputing some randomly chosen elements in the coefficients of each layer ($Z_1$, $Z_2$ and $Z$) with zeroes. In each iteration, once the coefficients are obtained by solving sub-problems P4, P5 and P6, we randomly drop some of their values to zeroes before they are used for updating the dictionary elements via P1, P2 and P3. The proportion of the elements to be dropped is user-defined. However, it must be noted that since we impose sparsity in the output, randomly dropping the coefficients from Z may not be sensible. It would be beneficial to restrict the droppings to the intermediate layers only.

---

[1] http://cs.nyu.edu/~wanli/dropc/

Similarly, we can incorporate DropConnect type regularization into the deep dictionary learning framework. Here, the dictionary atoms play the role of connections. Therefore, to drop connections, we can impute some randomly chosen elements in the dictionaries to be zeroes. After updating the dictionaries via P1, P2 and P3, we can impute some elements (randomly chosen) in the dictionaries to be zeroes to emulate DropConnect. Here too, the percentage of dropping need to be user-defined.

It must be noted that both DropConnect and DropOut type regularizations are only used till the pre-final iteration. In the final iteration, the obtained values of the dictionaries and the coefficients are not perturbed in any fashion.

## IV. EXPERIMENTAL RESULTS

We evaluate our proposed technique on two benchmark datasets.
- The Indian Pines dataset was collected by the Airborne Visible/Infrared Imaging Spectrometer in Northwestern Indiana, with a size of 145 × 145 pixels with a spatial resolution of 20 m per pixel and 10-nm spectral resolution over the range of 400–2500 nm. As is the usual protocol, the work uses 200 bands, after removing 20 bands affected by atmosphere absorption. There are 16 classes; the number of training and test samples is displayed in Table I.
- This Pavia university dataset is acquired by reflective optics system imaging spectrometer (ROSIS). The image is of 610 × 340 pixels covering the Engineering School at the University of Pavia, which was collected under the HySens project managed by the German Aerospace Agency (DLR). The ROSIS-03 sensor comprises 115 spectral channels ranging from 430 to 860 nm. In this dataset, 12 noisy channels have been removed and the remaining 103 spectral channels are investigated in this paper. The spatial resolution is 1.3 m per pixel. The available training samples of this data set cover nine classes of interests. Table II provides information about different classes and their corresponding training and test samples.

For both the datasets, the experimental protocol is borrowed from [6]; it is a very challenging protocol as the number of training samples is extremely limited. Some deep learning techniques (such as [1]) use about 90% of the data for training and validation. Prior studies based on support vector machines, random decision forests, sparse representation based classification and dictionary learning trained with only 10% of the data. Here, we use even less volume of training data. But this protocol reflects the real life scenario appropriately.

TABLE I
TRAINING AND TEST SAMPLES FOR INDIAN PINES

| Class | Training Samples | Test Samples |
|---|---|---|
| Alfalfa | 50 | 1384 |
| Corn-notill | 50 | 784 |
| Corn-min | 50 | 184 |
| Corn | 50 | 447 |
| Grass-pasture | 50 | 697 |
| Grass-trees | 50 | 439 |
| Grass-pasture mowed | 50 | 918 |
| Hay-windrowed | 50 | 2418 |
| Oats | 50 | 564 |
| Soybean-notill | 50 | 162 |
| Soybean-mintill | 50 | 1244 |
| Soybean-clean | 50 | 330 |
| Wheats | 50 | 45 |
| Woods | 15 | 39 |
| Buildings-glass-trees | 15 | 11 |
| Stone-steel-towers | 15 | 5 |
| Total | 695 | 9671 |

TABLE II
TRAINING AND TEST SAMPLES FOR PAVIA UNIVERSITY

| Class | Training Samples | Test Samples |
|---|---|---|
| Asphalt | 548 | 6631 |
| Meadows | 540 | 19649 |
| Gravel | 392 | 2099 |
| Trees | 524 | 3064 |
| Metal sheets | 265 | 1345 |
| Bare soil | 532 | 5029 |
| Bitumen | 375 | 1330 |
| Bricks | 514 | 3682 |
| Shadows | 231 | 947 |
| Total | 3921 | 42776 |

### A. Comparison with deep techniques

In this work we are going to compare with the latest deep learning-based techniques. The DBN based technique we compare against is [2]; the group-sparse formulation used there-in has shown excellent results for hyperspectral image classification. This technique is dubbed GBN in [2]. The benchmark for deep autoencoder (DAE) is the technique proposed in [5]. Following [6], we compare with their proposed method with recurrent neural network (RNN) based implementation. Of the PCA-Net variants, we compare against [7] that uses a non-linearity and fuses the spectral and spatial features in the final densely connected layer for classification. This technique [7] has been called NSS-Net. The final technique that we compare against is called deep fusion CNN (DFNN) [33]. It is a very recent work and to the best of our knowledge yields the best possible results. We also compare with the Robust deep dictionary learning formulation (RDDL) proposed in [16]. For all the methods mentioned here, we employ the best architecture and the corresponding parameter settings used in the papers.

For the proposed row sparse deep dictionary learning (RSDDL) formulation we use a three-layer architecture. For both the datasets the number of atoms in each layer are 100-50-25. The value of the sparsity parameter used here is $\lambda=0.1$ and the diversity parameter is $\mu=0.5$. We found that for both the parametric values, our method is fairly robust. Our algorithm also needs specification of the hyper-parameters. Usually in split Bregman techniques these need to be tuned. However, for our case, these hyper-parameters have special meanings. They imply the relative weights we give to each layer. Since there is no reason to favor one layer over the other, all of them have been fixed to unity. There is only one hyper-parameter we need to tune, i.e. $\gamma$ in (18). We kept its value to be 0.1. Since we are using the Split Bregman technique, the algorithm is robust to whatever value of γ we choose between 0.01 and 0.1. All the Bregman relaxation variables have been initialized to unity.

We found that DropOut does not help improve the results of our proposed technique. Instead, even with a small percentage of dropping (say 5%) the results deteriorate. But DropConnect regularization improves our results slightly. The best results are obtained for 10% dropping; it improves the results by about 1.2 % on average. Therefore in this work, we do not use DropOut; we only use DropConnect with 10% dropping.

Note that all the parametric values and the user defined dropping percentages used in this work have been tuned on a third dataset, namely the Salinas dataset. We tuned the parameters using a greedy L-curve method [34]. Here the parameter λ is first tuned (using L-curve) by fixing μ to Zero. Once λ is tuned, it is fixed at that value and then μ is tuned by the L-curve method. These values have not been tuned on the Indian Pines and the Pavia University datasets.

The inputs to our proposed RSDDL are spatio-spectral features. These have been derived in the way proposed by [1, 4], i.e. a window of size 4x4 is captured and all the bands within the window are taken. The dimensionality is reduced to 200 by principal component analysis. This is used as input to our proposed method.

Evaluation is carried out by the standard measures of Average Accuracy (AA), Overall Accuracy (OA) and Kappa (K) coefficient. The definitions for these terms are given below.
- Average Accuracy – This measure is the average value of the classification accuracies of all classes.
- Overall Accuracy – This index shows the number of hyperspectral pixels that are classified correctly, divided by the number of test samples.
- Kappa Coefficient – This metric is a statistical measurement of agreement between the final classification map and the ground-truth map. It is the percentage agreement corrected by the level of agreement that could be expected due to chance alone. It is generally thought to be a more robust measure than a simple percent agreement calculation, since it takes into account the agreement occurring by chance

The numerical results (in terms of the aforesaid indices) for both the Indian Pines and Pavia University datasets are shown in Table III.

TABLE III
COMPARISON WITH STATE-OF-THE-ART DEEP LEARNING TECHNIQUES

| Dataset | Metric | GBN [2] | DAE [5] | RNN [6] | NSS-Net [7] | DFNN [12] | RDDL [16] | Proposed-0 | Proposed-1 |
|---|---|---|---|---|---|---|---|---|---|
| Pavia | OA | 86.77 | 88.36 | 88.87 | 89.55 | 88.50 | 88.52 | **90.91** | 89.30 |
|  | AA | 85.31 | 86.73 | 86.43 | 89.03 | 86.37 | 87.38 | **90.22** | 88.62 |
|  | Kappa | 0.79 | 0.80 | 0.80 | 0.79 | 0.79 | 0.80 | **0.86** | 0.84 |
| Indian Pines | OA | 85.42 | 89.02 | 88.59 | 88.82 | 86.34 | 88.76 | **90.76** | 89.14 |
|  | AA | 86.31 | 85.86 | 85.36 | 86.48 | 85.09 | 85.93 | **88.49** | 87.68 |
|  | Kappa | 0.74 | 0.75 | 0.73 | 0.75 | 0.74 | 0.75 | **0.78** | 0.77 |

*Proposed-0: $l_0$-norm between training and test features; Proposed-1: $l_1$-norm between training and test samples

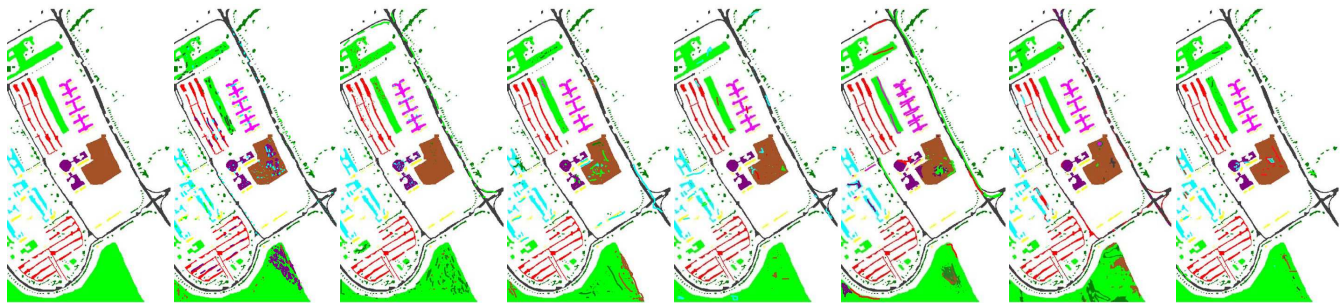

**Fig.** 3. Pavia University. Left to Right – Groundtruth, GBN, DAE, RNN, NSS-Net, DFNN, RDDL and Proposed-0.

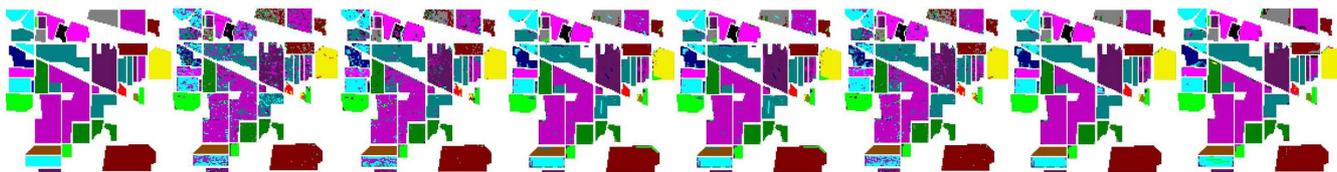

**Fig.** 4. Indian Pines. Left to Right – Groundtruth, GBN, DAE, RNN, NSS-Net, DFNN, RDDL and Proposed-0.

In terms of numbers, we see that both versions of our proposed method clearly outperform others. However, the $l_0$-norm outperforms the $l_1$-norm; even though the later is supposed to be theoretically sounder. This may be owing to two reasons. First, the Hamming distances (for $l_0$-norm) are never tied in practice. Second, the $l_1$-norm does not enforce exact sparsity and hence obfuscate the results.

The DAE, NSS-Net, RNN and RDDL perform almost the same and is worse than ours. The GBN based formulation is slightly worse than the aforesaid techniques. The CNN formulation performs at par with other existing methods. We tried our best to implement the CNN based technique, however with the best of our efforts we were not able to achieve the accuracy reported in there for the Indian Pines dataset. This may be owing to the randomness in seeding the CNN or in selecting the training samples.

For visual evaluation we have shown the results in Fig.s 3 and 4 for the Pavia and the Indian Pines dataset respectively. Since, the Proposed-0 approach is the best of our two approaches, we show results only from it. Visual evaluation corroborates the numerical results.

In order to test the statistical significance of our results, we have used the McNemar's test (also carried out in [12]). The results are shown in Table IV. It can be clearly seen that our Proposed-0 method is indeed significantly better than the others.

TABLE IV
STATISTICAL SIGNIFICANCE FROM STANDARDIZED MCNEMAR'S TEST

| Proposed Vs. | Pavia University Z / significant | Indian Pines Z / significant |
|---|---|---|
| GBN | 24.33 / yes | 29.12 / yes |
| DAE | 20.19 / yes | 17.60 / yes |
| RNN | 20.67 / yes | 21.09 / yes |
| NSS-Net | 15.70 / yes | 17.61 / yes |
| DFNN | 26.71 / yes | 32.64 / yes |
| RDDL | 19.56 / yes | 18.97 / yes |

All the experiments were conducted on an Intel Xeon E3-1246 CPU at 3.5 GHz with 32-GB RAM using a MATLAB platform. The training and testing times for all the aforesaid techniques are given in Table V.

TABLE V
TRAINING TIME IN SECONDS

| Technique | Pavia | | Indian Pines | |
|---|---|---|---|---|
| | Training | Testing | Training | Testing |
| GBN | 14443 | 41 | 4591 | 10 |
| DAE | 8291 | 43 | 2307 | 11 |
| RNN | 4095 | 90 | 1002 | 21 |
| NSS-Net | 1906 | 102 | 1523 | 28 |
| DFNN | 12409 | 97 | 8002 | 23 |
| RDDL | 70 | 68 | 45 | 16 |
| Proposed-0 | 102 | 95 | 87 | 22 |
| Proposed-1 | 102 | 105 | 87 | 25 |

Note that for training, there is no difference between our two proposed approaches, since it only pertains to testing. In terms of training, one can see, RDDL is the fastest; this is expected because it is a greedy learning approach. Our method takes slightly more time but is several orders of magnitude faster than other techniques.

For testing, we find that GBN and DAE are the fastest; this is because they only require a few matrix vector products during testing. RSDDL is slower compared to these, because one needs to solve inverse problems (albeit in a closed form); owing to the non-linearity the inverses cannot be pre-computed. Our proposed method are relatively slow owing to the necessity of solving involved optimization problems during testing. We find that the $l_0$-norm based solution is faster compared to the $l_1$-norm (the reason has been explained before). The CNN based solutions are slower than our best performing overall technique ($l_0$-norm), but the RNN based solution is slightly faster than ours.

B. Comparison with shallow techniques

Since deep learning techniques are data hungry and do not perform well in limited training data scenarios, we compared with traditional (shallow) machine learning techniques as well. Label consistent KSVD (LC-KSVD) [35] was proposed as a generic classification algorithm based on the popular dictionary learning framework; later it was adopted for hyperspectral image classification [36]. The results were LC-KSVD [36] were better than vanilla implementations of other machine learning algorithms, so use it as a benchmark.

Another technique that is compared against is the rotation based support vector machine RO-SVM [37]; this was specifically designed to function in the limited training data scenario.

The third method compared against is based on the extreme learning machine (ELM) formulation [38]. This is a state-of-the-art technique known for its very short training time.

The experimental protocol remains the same. For each of the aforesaid methods, we have taken the best configuration specified in the papers. However, we have not used fusion approaches used in some of the studies as a post processing step to boost the accuracy.

TABLE VI
COMPARISON WITH STATE-OF-THE-ART SHALLOW TECHNIQUES

| Dataset | Metric | RO-SVM | LC-KSVD | ELM | Proposed |
|---|---|---|---|---|---|
| Pavia | OA | 75.99 | 74.02 | 89.10 | **90.91** |
| | AA | 85.77 | 73.42 | 89.08 | **90.22** |
| | Kappa | 0.73 | 0.70 | 0.77 | **0.86** |
| Indian Pines | OA | 80.14 | 75.02 | 88.01 | **90.76** |
| | AA | 88.84 | 72.32 | 86.23 | **88.49** |
| | Kappa | 0.81 | 0.70 | 0.75 | **0.78** |

Note that one cannot compare the results obtained here and those reported in the previous papers for two reasons – 1. The experimental protocols are different; and, 2. There is no post-processing step used in our experiments.

The results show that of our method (with $l_0$-norm) still yields the best results. Results from RO-SVM and LC-KSVD are much worse than ours and all the deep learning techniques compared against. Results from ELM are comparable with other deep learning techniques, but is still worse than ours. As before, we perform McNemar's test to show the statistical significance (improvement) of our method.

TABLE VII
STATISTICAL SIGNIFICANCE FROM STANDARDIZED MCNEMAR'S TEST

| Proposed Vs. | Pavia University Z / significant | Indian Pines Z / significant |
|---|---|---|
| RO-SVM | 35.58 / yes | 29.36 / yes |
| LC-KSVD | 32.77 / yes | 32.08 / yes |
| ELM | 18.13 / yes | 17.42 / yes |

In the following table we show the training and testing times. ELM is the fastest; this was supposed to be given its design (closed form solution). SVM is the slowest in terms of training; its testing times are comparable with other dictionary learning based methods. It is interesting to note that even though LC-KSVD is a shallow method, it is slower than ours; this because of the requirement of solving the dictionary learning by the inefficient KSVD algorithm.

TABLE VIII
TRAINING AND TESTING TIMES IN SECONDS

| Technique | Pavia | | Indian Pines | |
|---|---|---|---|---|
| | Training | Testing | Training | Testing |
| RO-SVM | 1002 | 100 | 255 | 58 |
| LC-KSVD | 119 | 97 | 107 | 27 |
| ELM | 31 | 40 | 8 | 12 |
| Proposed | 102 | 95 | 87 | 22 |

*C. Effect of Stochastic Regularization*

In this sub-section we have elaborated on the different dropping rates. The kappa coefficients are shown in Table IX. The columns of the tables should be read independently, i.e. the DropOut and the DropConnect have been used separately and not combined with each other. Since the Proposed-o method is better than the other approach, we show results from the former only.

TABLE IX
VARIATION OF KAPPA WITH DROPPING RATES

| %age of dropping | Pavia | | Indian Pines | |
|---|---|---|---|---|
| | DropOut | DropConnect | DropOut | DropConnect |
| 0 | 0.82 | 0.82 | 0.73 | 0.73 |
| 2.5 | 0.81 | 0.83 | 0.72 | 0.75 |
| 5 | 0.78 | 0.84 | 0.70 | 0.77 |
| 10 | 0.74 | 0.86 | 0.66 | 0.78 |
| 15 | 0.68 | 0.82 | 0.62 | 0.73 |

The results show that with DropOut, the results deteriorate very fast. This is the reason, we have not used it in our main experiments. With DropConnect, the results improve till about 10%, but with further dropping, the result deteriorates.

We believe that DropOut degrades performance, both our deterministic sparsity promoting regularization ($l_0$-norm) and DropOut (random zero-ing out of features) negates each other. The $l_0$-norm enforces sparsity and selects very few coefficients; further dropping after having a sparse output tends to lose information and hence deteriorates.

DropOut prevents coadaptation of nodes. To a certain extent it helps barring overfitting; but when too many are dropped, the network lacks the capacity to model the problem. Hence the results deteriorate (after 10%).

## V. CONCLUSION

In this work we propose a new technique for hyperspectral image classification based on the deep dictionary learning framework. We show that the proposed technique can yield good results even with extremely few training samples. Comparison has been carried out with state-of-the-art deep learning based methods published in the last year. Our method outperforms them all by a statistically significant margin. We have also carried out comparison with several traditional (shallow) machine learning approaches particularly tailored for the said problem published in the last few years; in terms of accuracy we excel over these as well.

In terms of technique there are several contributions of this work. First, ours is the only work that can solve all the layers of deep dictionary learning problem in a joint fashion; instead of solving one layer at a time in a greedy fashion (as is done in prior studies), we learn all the layers jointly. Second, we propose a new discriminative penalty. Third, we introduce stochastic regularization techniques in lines with DropConnect and DropOut.

Although used in this work for the purpose of hyperspectral image classification; the method proposed here is fundamental and can be applied to a variety of problems.

ACKNOWLEDGEMENT

This work is supported by the Infosys Center for Artificial Intelligence @ IIIT Delhi and by the Indo-French CEFIPRA grant DST-CNRS-2016-02.